\documentclass[english,aps,prl,twocolumn, superscriptaddress,showpacs, amsmath]{revtex4-1}

\usepackage[T1]{fontenc}
\usepackage[latin9]{inputenc}
\usepackage{color}
\usepackage{units}
\usepackage{amssymb}
\usepackage{graphicx}
\usepackage{esint}
\usepackage{bm}
\usepackage{natbib}
\usepackage{ulem}
\usepackage[colorlinks=true, citecolor=red]{hyperref}
\setcitestyle{journalcolor= blue}

\newcommand{\vl}{$v^l$ }
\newcommand{\vnl}{$v^{nl}$ }
\usepackage{graphicx}
\usepackage{dcolumn}
\usepackage{bm}

\usepackage{babel}

\begin{document}

\title{Robust local and non-local transport in the Topological Kondo Insulator SmB$_{6}$ in the presence of high magnetic field}
\author{Sangram Biswas}

\affiliation{Department of Physics, Indian Institute of Science, Bangalore 560012, India}

\author{Ramya Nagarajan}

\affiliation{Department of Physics, Indian Institute of Science, Bangalore 560012, India}

\author{Suman Sarkar}

\affiliation{Department of Physics, Indian Institute of Science, Bangalore 560012, India}

\author{Kazi Rafsanjani Amin}

\affiliation{Department of Physics, Indian Institute of Science, Bangalore 560012, India}

\author{M. Ciomaga Hatnean}
\affiliation{Department of Physics, University of Warwick, Coventry, CV4 7AL, UK.}

\author{S. Tewari}
\affiliation{Department of Physics and Astronomy, Clemson University, Clemson, SC 29634, USA.}

\author{G. Balakrishnan}
\affiliation{Department of Physics, University of Warwick, Coventry, CV4 7AL, UK.}

\author{Aveek Bid}
\email{aveek.bid@physics.iisc.ernet.in}
\affiliation{Department of Physics, Indian Institute of Science, Bangalore 560012, India}

\begin{abstract}

SmB$_6$ has been predicted to be a Kondo Topological Insulator with topologically protected conducting surface states. We have studied quantitatively the electrical transport through surface states in high quality single crystals of SmB$_6$. We observe a large non-local surface signal at temperatures lower than the bulk Kondo gap scale.  Measurements and finite element simulations allow us to distinguish unambiguously between the contributions from different transport channels. In contrast to general expectations, the electrical transport properties of the surface channels was found to be insensitive to high magnetic fields. Local and non-local magnetoresistance measurements allowed us to identify definite signatures of helical spin states and strong inter-band scattering at the surface. 
 
\end{abstract}

\maketitle

Topological insulators (TI) are a new class of materials where gapped bulk states coexist with gapless surface states having linear energy-momentum dispersion relation protected by time reversal symmetry~\cite{hassanrmp, moore, ando, zhangrmp}. These surface states are the consequence of the closing of the bulk gap at an interface as a result of the change of the non-trivial topology of the Hilbert space spanned by the wave function describing the insulator to a trivial one at the boundary of the material. The large spin-orbit coupling of the bulk material implies that Dirac fermions at the interface have helical spin polarization - the spin degeneracy of the Dirac fermions is lifted and the spin becomes transversely locked to the crystal momentum~\cite{PhysRevLett.95.226801}.  This leads to interesting possibilities, chief among them being the existence of majorana modes in proximity-induced superconducting state~\cite{PhysRevLett.100.096407}. This, and the obvious possible applications in spintronics has led the search for new robust TI materials.

The heavy fermion SmB$_6$ has recently been predicted to be a topological Kondo insulator (TKI)~\cite{dzero2010topological,lu2013correlated}.  At temperatures below few tens of Kelvin a bulk energy gap opens up in this material due to hybridization of the conducting electrons in the 5d band and the localized moments in the 4f band leading to the appearance of a Kondo insulator state. In electrical transport measurements this shows up as a rapid increase in the measured resistance as the material is cooled down below about 40~K. The presence of the bulk gap in SmB$_6$ down to atleast 2~K has also been seen through capacitance measurements~\cite{kim2012limit} and point-contact spectroscopy~\cite{zhang2013hybridization}. Below 4~K the resistance was seen to saturate which could not be explained by the standard theory of Kondo insulators. Recently, it has been proposed that this low temperature saturation of the resistance was due to the appearance of topologically protected surface states in the Kondo gap. Existence of this surface state has been confirmed through resistivity measurements ~\cite{kim2014topological,kim2013surface,wolgast2013low}, non-local transport ~\cite{wolgast2013low,kim2013surface}, ARPES  ~\cite{jiang2013observation,xu2014exotic,xu2014direct}, magneto-conductance measurements ~\cite{kim2014topological,kim2013surface} and point contact spectroscopy~\cite{zhang2013hybridization}. An important implication of the existence of topologically protected surface states is the existence of spin-momentum locking on the surface~\cite{PhysRevLett.95.226801}. The cleanest signatures of this phenomenon comes out from spin-resolved ARPES measurements - till date there is no convincing demonstration of this for the case of SmB$_6$. An alternate way of addressing this issue is through local and non-local magneto-transport studies which can probe the existence of spin helical surface states~\cite{PhysRevLett.108.036805, PhysRevB.90.165140, PhysRevLett.106.166805, 2013arXiv1307.4133T}. 

\begin{figure}[!]
\begin{center}
\includegraphics[width=0.5\textwidth]{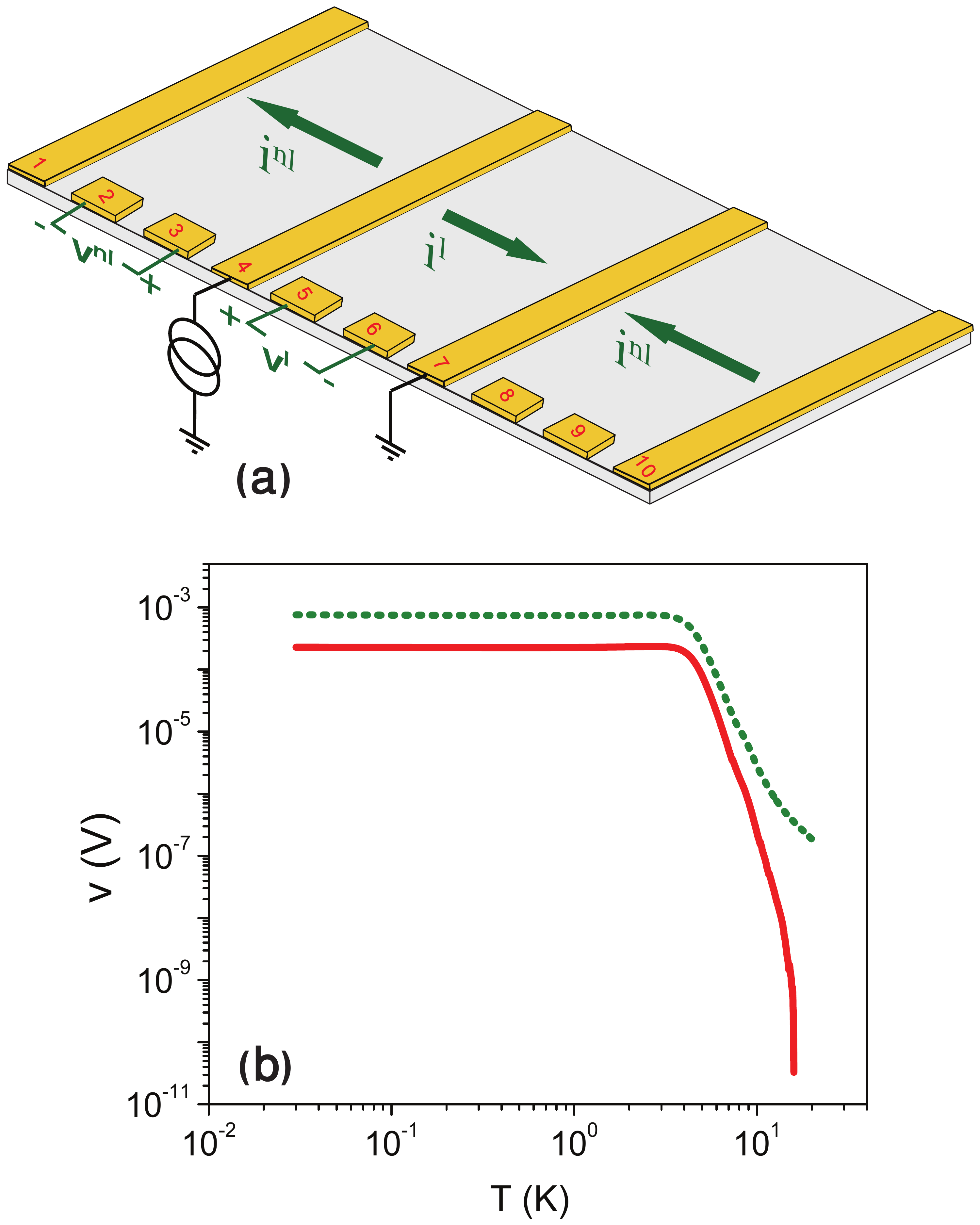}
\small{\caption{(color online) Local and non-local transport in SmB$_6$. (a) The figure shows the  local and nonlocal measurement schemes for configureation~A, arrows denote the direction of current flow. Contact pads numbered 1, 4, 7 and 10 were used as current and ground probes - these probes had dimensions $600 \times 100 \mu$m$^2$. Rest of the probes had dimensions $100 \times 100 \mu$m$^{2}$ and were used as voltage probes. (b) \vl (olive dotted line) and \vnl (red solid line) measured in configuration~A, the local voltage was measured between contact pad 5 and 6; 3 and 2 were used for nonlocal voltage measurements. \label{fig:rt}}}
\end{center}
\end{figure}

We have studied in detail local and non-local magneto-transport on high quality single crystal samples of SmB$_6$. Our studies allow us to identify definite signatures of helical spin states and strong inter-band scattering at the surface.  The samples were prepared by the floating zone technique using a high power xenon arc lamp image furnace~\cite{balakrishnan}. Electrical contacts were defined on the (110) surface by standard electron beam lithography followed by Cr/Au deposition. Before lithography the sample surface was cleaned using concentrated HCl and then mirror polished to get rid of any surface contaminants. Defining the contacts lithographically enabled us to determine precisely the distance between the electrical probes which was essential for the analysis presented in this letter. Electrical measurements were done using many different contact configurations on two different samples to unambiguously probe the local and non-local transport in the system - in the rest of the letter we primarily discuss the data obtained in configuration~A, described in figure~\ref{fig:rt}(a). Electrical measurements were performed using standard low frequency ac measurement techniques by current biasing the sample (10~nA-100~nA at $T<2$~K, upto 10$\mu$~A at higher temperatures) in the temperature range 10 mK to 300 K and up to 16~T magnetic field. \\

Figures~\ref{fig:rt}(b) shows the measured local \vl and non-local \vnl voltages as a function of temperature measured at zero magnetic field. The local  voltage \vl increases by five orders of magnitude as the device is cooled down from room temperature to below 1~K attesting to the high quality of the sample. The value of the bulk band gap $E_g$ extracted from \vl \textit{vs} temperature measurements increases from 3.5~meV at 20~K to 5.5~meV at 9~K - this increase of the bulk kondo gap with decreasing temperature is consistent with previous reports~\cite{zhang2013hybridization, allen1979large}. 

A significant non-local voltage \vnl appears below 15~K and increases rapidly by several orders of magnitude before saturating at temperatures below 3~K. The temperature at which the non-local signal appears corresponds to the energy scale of the many body Kondo gap in this system.  This appearance of the non-local signal concomitantly with the opening of the bulk Kondo gap strongly indicates that the 2-dimensional surface conduction channels in SmB$_6$ emerge from the bulk topological Kondo insulator state and are not the result of accidental surface states. (This was also verified by preparing the surface in different ways by varying the surface polishing method and duration of exposure to the ambient - the results were all quantitatively consistent with each other.)  The sharp increase in \vnl with decreasing temperature observed in our measurements can either be due to an increase in the resistance of the surface channel, $R_s$ or an increase in the fraction of total current flowing through the surface channel. As shown later in this letter, the resistance of the surface channel is almost independent of temperature and the increase in measured \vnl is because of an increase in surface current with decreasing temperature.

\begin{figure}[!]
\begin{center}
\includegraphics[width=0.45\textwidth]{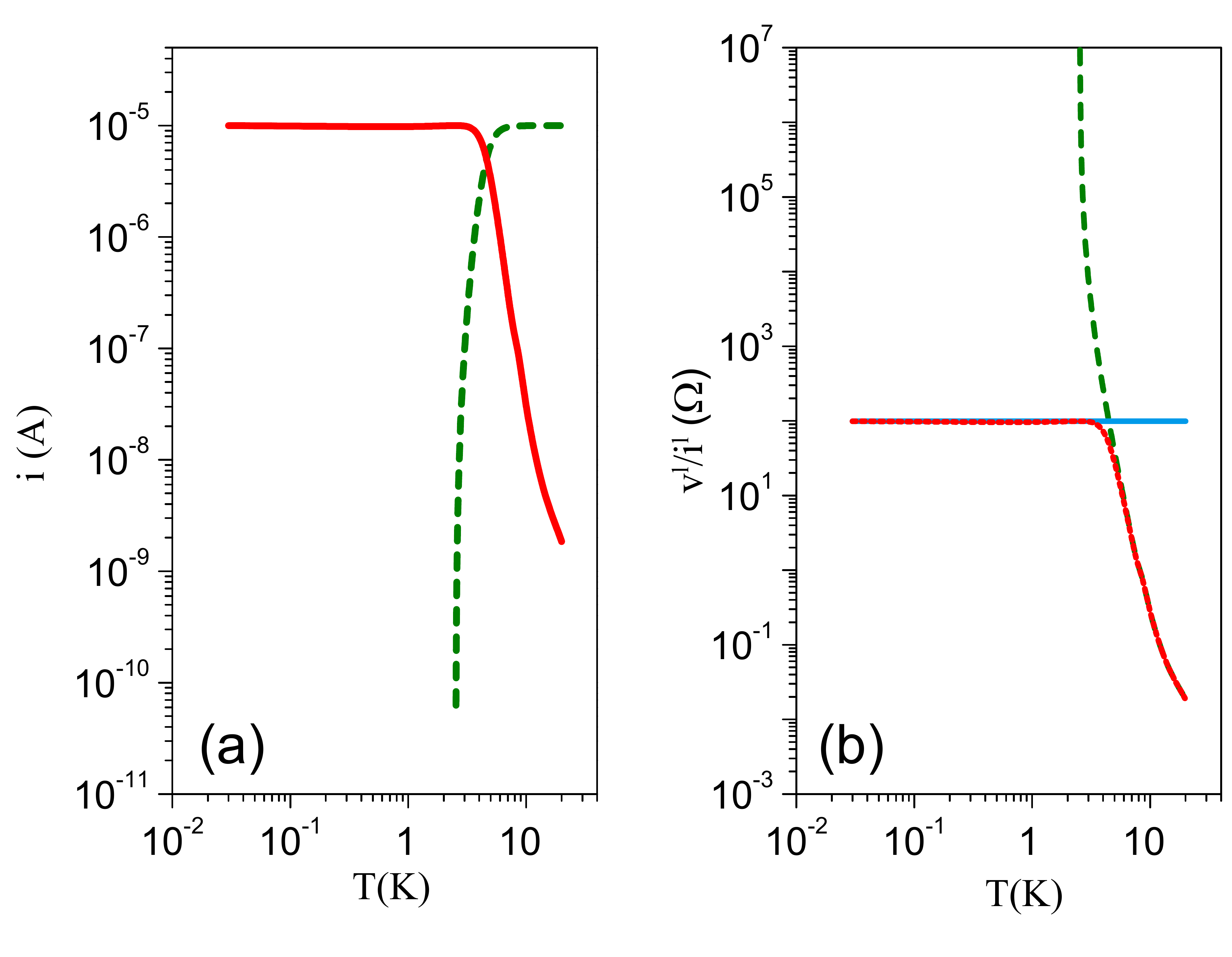}
\small{\caption{ (a) Amount of current flowing through the bulk states (olive dotted line) and surface states (red solid line) as a function of temperature. Below 3K almost the entire current flows through the surface states as the resistance of the bulk diverges exponentially.  (b) Green dashed line and cyan solid line show respectively the values of $R_b$ and $R_s$ as function of temperature as extracted from our analysis, the red dotted line shows the experimentally measured $v^l/i^l$ for the device. \label{fig:isib}}}
\end{center}
\end{figure}

The total current driven through the sample gets distributed into three channels:

\begin{eqnarray}
i_{0}= i_{b}(T)+i^{l}_{s}(T)+i^{nl}_{s}(T).
\label{eqn:current}
\end{eqnarray}

\noindent Here $i_{0}$ is the total applied current bias, $i_{b}$(T) is the current flowing through the bulk of the sample, $i^{l}_{s}$(T) is the local surface current and $i^{nl}_{s}$(T) is the non-local surface current.  Equation~\ref{eqn:current} combined with the fact that the bulk resistance is activated down to the lowest temperature measured~\cite{zhang2013hybridization, kim2012limit} allows us to extract the values of bulk and the surface current components over the entire temperature range - the results are plotted in figure~\ref{fig:isib}(a). The results can be understood using the following argument: At high temperatures (above 15~K) current flows only through the bulk channel as the bulk resistance is very low in this temperature range. As the temperature decreases, the resistance of the bulk channel $R_b$ diverges exponentially due to the opening up of the bulk energy gap. $R_s$ on the other hand remains almost constant with temperature. Consequently, the proportion of surface current increases rapidly with decreasing temperature and at very low temperatures almost the entire current flows through the surface channels. The exact division of the surface current into the local $i^{l}_{s}$(T)  and the non-local $i^{nl}_{s}$(T) components depends on  the relative positions of the electrical contacts on the device  and on the anisotropy of the sheet resistance tensor.  Figure~\ref{fig:isib}(b) shows a plot of the resistance of the surface states $R_s$ and the bulk resistance $R_b$ of the device extracted from our analysis  along with the measured $v^l/i^l$. The agreement of our calculated value with the experimental result supports the picture that $R_s$ is almost constant throughout the temperature range below 15~K. However, a very small variation of $R_s$ with temperature is found when temperature in lowered below 3~K which hints at richer physics underlying the surface states.

\begin{figure}[!]
\begin{center}
\includegraphics[width=0.45\textwidth]{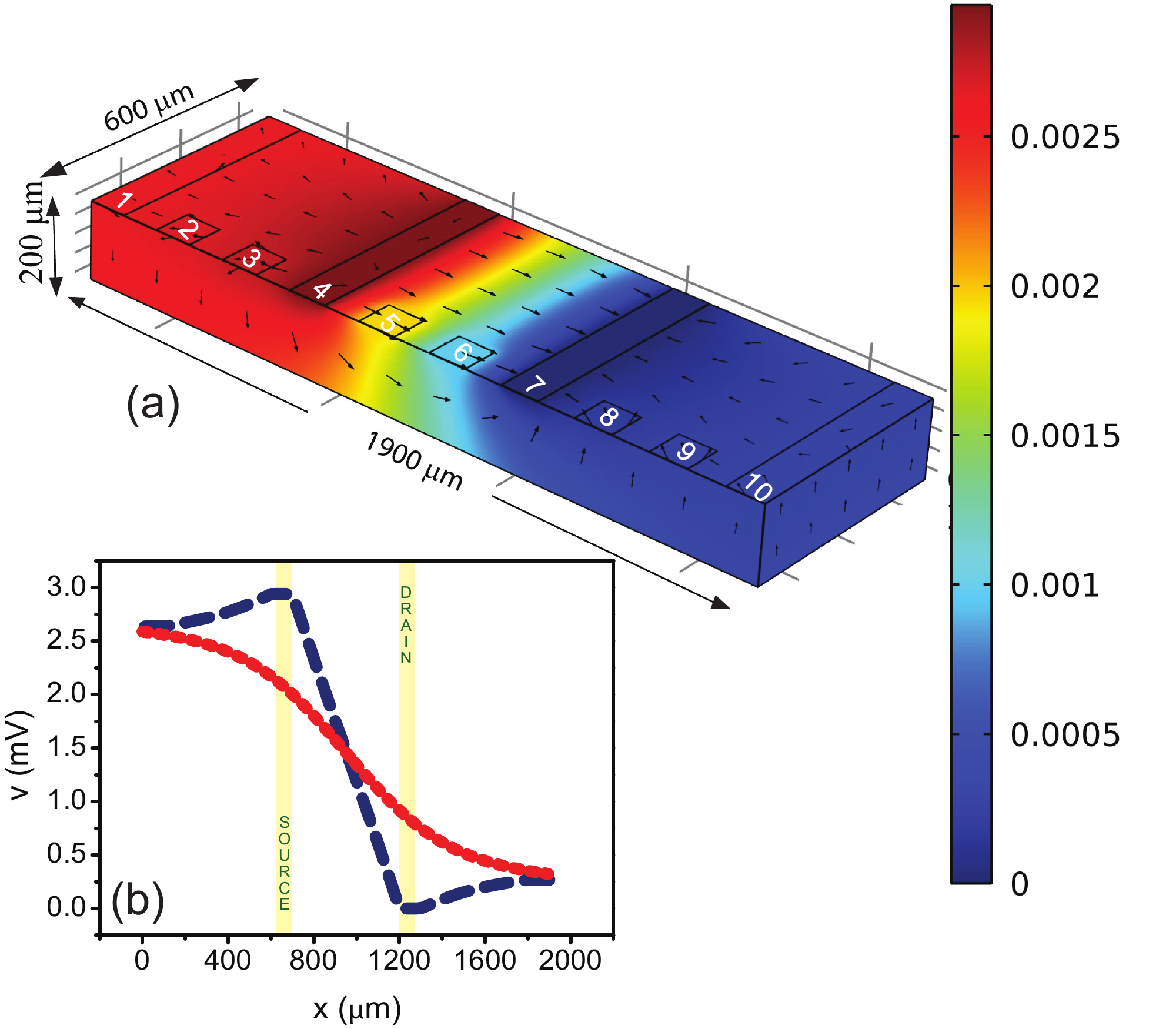}
\small{\caption{ (a) Electric field profile obtained from the finite element simulations. The potential at contact~4 was set at 2.94~mV and contact pad~7 was grounded. The rest of the pads were left floating. The arrows denote the direction of the tangential surface current density. (b) Plot of the line profile of the electric potential on the surface extracted from the simulations as a function of the distance $x$ from the left edge of the device. The blue dashed line shows the potential profile on the top surface while the red dotted line shows the potential profile for the bottom surface. The yellow shaded areas mark the position of the source and drain contact pads on the top surface.\label{fig:sim_A}}}
\end{center}
\end{figure}

\begin{figure}[!]
\begin{center}
\includegraphics[width=0.45\textwidth]{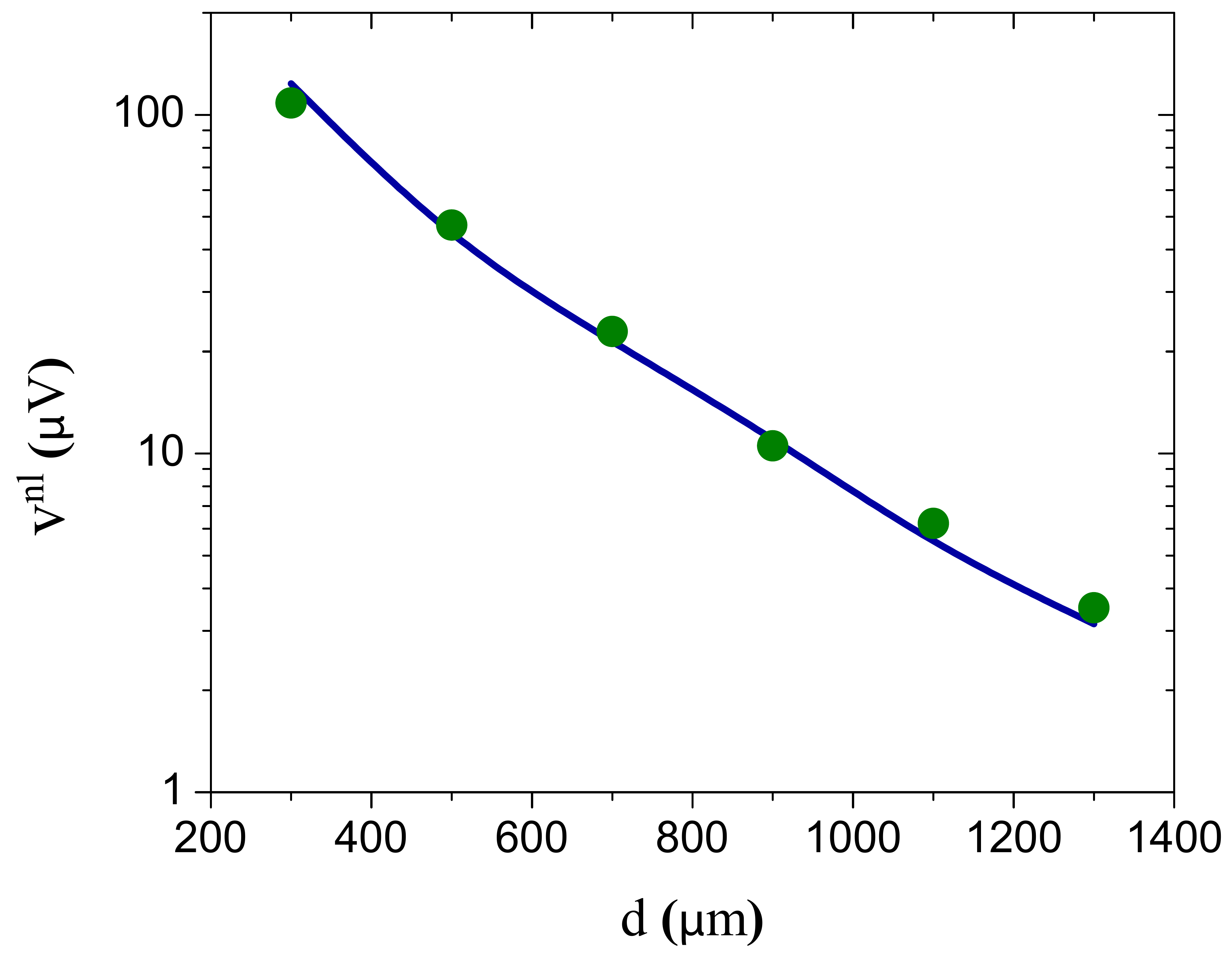}
\small{\caption{Olive filled circles show the measured \vnl for voltage probes separated by 200$\mu$m and located at an average distance $d$ away from current probe. In this configuration current was sourced from contact pad 9 and pad 10 was grounded. Blue dotted line is obtained from the simulations for the same contact geometry.   \label{fig:nlsim}}}
\end{center}
\end{figure}

To get a quantitative understanding of the local and non-local surface transport in our device we performed extensive simulations based on finite-element method using COMSOL Multiphysics modelling software~\cite{comsol}. The simulations were performed for the exact dimensions of our device. The system was modelled as an ohmic conducting surface with an insulating bulk. An example of the surface current densities and the electric field profile obtained for configuration~A is shown in figure~\ref{fig:sim_A}(a). (see supplementary section for more details on the simulation technique and results). The line profile of the electric potential on the surface was extracted from the simulations and is plotted in figure~\ref{fig:sim_A}(b). The blue dashed line shows the potential profile on the top surface while the red dotted line shows the potential profile for the bottom surface. The yellow shaded areas mark the position of the source and drain contact pads on the top surface. Using this model, we calculated the dependence of the non-local surface potential at various distance from source contact $d$. The analysis of the non-local potential was complicated by the fringing of the electrostatic equipotential lines as one moves away from the source contact. To quantify the effect of this fringing on the non-local transport, we measured \vnl between equally spaced voltage probes located at different distances $d$ from the source probe. Figure~\ref{fig:nlsim} shows the measured and calculated (from simulations) \vnl data. The data shows that the non-local voltage measured between equally spaced voltage probes falls off rapidly as the distance between the voltage probes and the source probe increases. The excellent agreement between the measurement and simulation results lends credence to our model that transport in the ultra-low temperature range is through surface states only.

\begin{figure}[!]
\begin{center}
\includegraphics[width=0.45\textwidth]{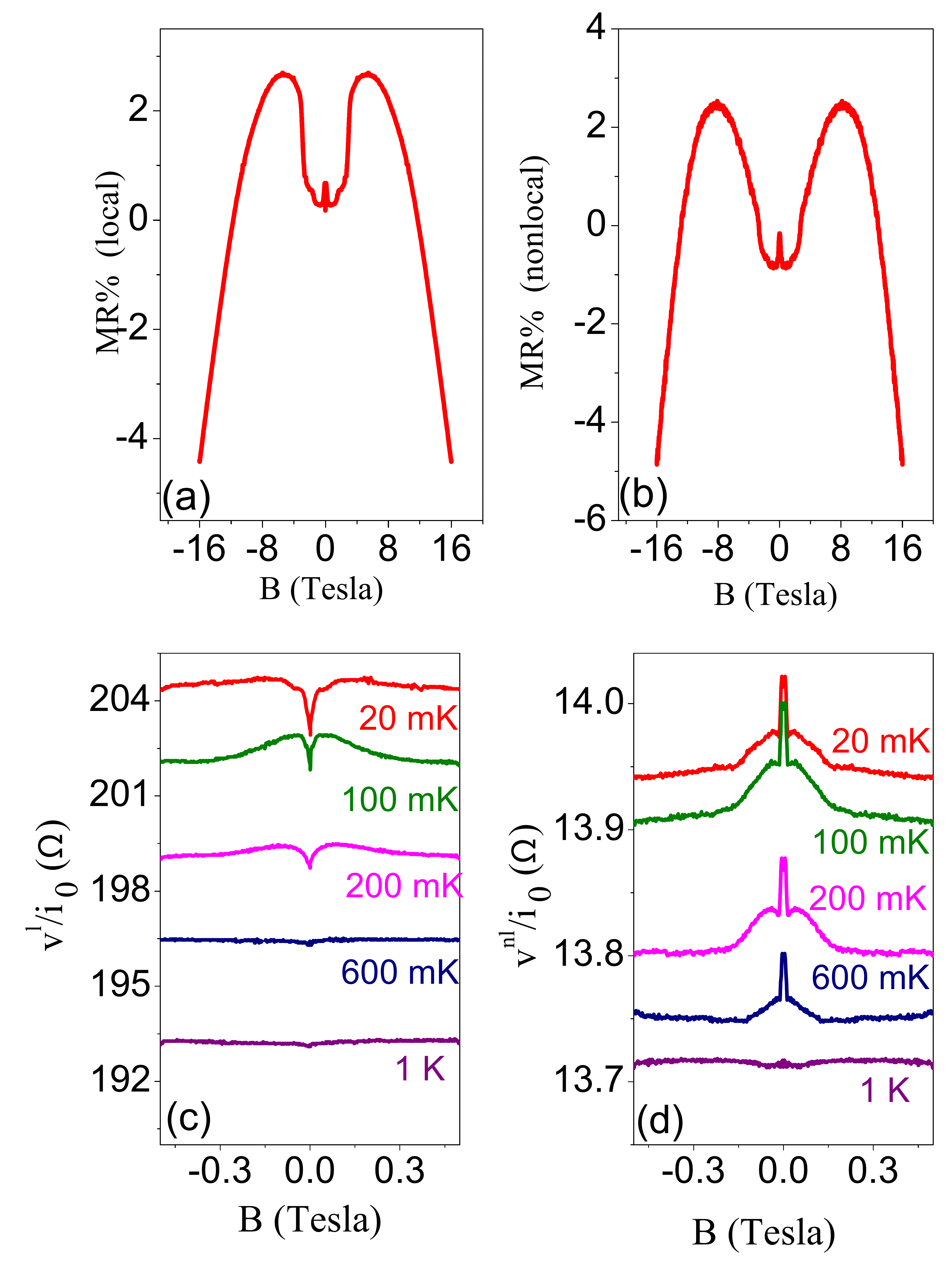}
\small{\caption{Magnetoresistance of the device (a) Local MR measured in the range -16T to 16T. (b) non-local MR measured in the range -16T to 16T. In both cases the measuremets were done at 20mK. (c) Low field local MR showing weak anti-localization. (d) Low field non-local MR. \label{fig:mr}}}
\end{center}
\end{figure}

To probe the effect of magnetic field on the surface states in SmB$_6$, we carried out local and non-local magneto-transport  measurements in magnetic fields upto 16~T. If the system at low temperatures indeed goes into a topological Kondo insulator state, then conventional wisdom suggests that the surface states should be quenched in the presence of a magnetic field. Indeed it has been shown in a previous work that the presence of magnetic impurities quenches the  surface states in SmB$_6$ ~\cite{kim2014topological}. In figure~\ref{fig:mr}(a) we plot the local magnetoresistance (MR) measured at 20~mK while the non-local MR is plotted in \ref{fig:mr}(b). In both cases the MR is very small - of the order of a few percentage even at 16~T magnetic field.  This is in contrast to what we expect for TI surface states in the presence of a time reversal breaking field.  On the other hand it has been recently suggested~\cite{PhysRevLett.95.136602, PhysRevLett.97.036808, PhysRevB.80.125327, PhysRevLett.107.066602} and experimentally demonstrated~\cite{du2013observation} that robust topologically protected spin helical states can indeed exist even under the conditions of broken time reversal symmetry. This magnetic field insensitivity of topological surface states in two-dimensional topological insulators has been ascribed to a spin-Chern topological invariant. We note that the conduction through the surface states of time reversal invariant topological insulators is also expected to be insensitive to time reversal symmetry (TR) breaking perturbations  if the Fermi energy of the surface Dirac cones is not at zero energy (which is where a gap is expected to open with TR breaking) and if the surface is relatively free of impurities. \\

The sharp dip near zero magnetic field in the local MR, seen more clearly in the zoomed in graph in figure~\ref{fig:mr}(c), is due to weak anti-localization effect (WAL) which is expected in the case of systems with large spin-orbit coupling. The magnitude of the dip decreases with increasing temperature eventually vanishing above 2~K. This can be understood as follows:  with increasing temperature the inelastic scattering rate of the charge carriers increase which eventually destroys phase coherence of the conducting electrons  essential for WAL~\cite{PhysRevB.28.2914, PhysRevB.68.085413}. The low field local magneto-condutance $\sigma(B)$ at different temperatures was fitted to the Hikami-Larkin-Nagaoka (HLN) equation in the presence of large spin-orbit coupling~\cite{hikami1980spin}: 

\begin{eqnarray}
\sigma(B)-\sigma(0) = \alpha\frac{e^2}{2\pi^2\hbar} \Bigg(ln\Bigg(\frac{B_{\phi}}{B}-\psi(\frac{1}{2}+\frac{B_{\phi}}{B})\Bigg)\Bigg)
\end{eqnarray}

\noindent where $B_{\phi}$ is related to the phase coherence length $l_{\phi}$ as $B_{\phi}=h/(4el_{\phi}^2)$. The fits yield $\alpha = 0.5$ in the temperature range 1~K > T > 0.1~K which is the value expected for 2-dimensional TI having a single coherent conducting channel. As the temperature decreases, the value of $\alpha$ steeply increases and saturates at temperatures below 50~mK to a value of about 1 which is the value expected for two independent topological coherent channels. Although this doubling of the parameter $\alpha$ with decreasing temperature has been reported recently in the case of the TI material Bi$_2$Te$_3$~\cite{PhysRevB.87.035122}, a proper understanding of this effect must await further theoretical investigations. In the absence of such detailed theoretical modelling we conjecture that the saturation of the value of $\alpha$ to about 1 at the lowest temperatures indicates either the dominance of two surface Dirac bands from amongst  the expected three bands or strong inter-band scattering so that there are effectively only two conduction channels. The value of $l_{\phi}$ obtained from the fits to the low-field MR data   was $\approx$ 700~nm at the lowest temperature. $l_{\phi}$ decreases with increasing temperature as a power law $l_{\phi} \propto T^{-p/2}$ with $p \approx 1$ over the entire temperature range measured suggesting that the major source of scattering is electron-electron interactions.

The low field behaviour of the non-local MR, shown in figure~\ref{fig:mr}(d) is not so clear as it is in the case of local MR. A possible reason could be that the local magneto-transport is confined only to the top surface of the device and hence is decided by the dispersion relation of only the [110] surface. The non-local current, on the other hand flows through multiple (not only the top and bottom, but also the side) surfaces of differing crystallographic orientation and hence differing dispersion relations. This makes the low-field non-local MR very dependent on the exact electric and magnetic field profiles and hence quite difficult to tackle analytically. Understanding the non-local magneto-transport in these systems will probably require significant theoretical investigations. 

To conclude, we have studied the temperature and magnetic field dependence of the non-local transport in the topological Kondo Insulator SmB$_6$. In contrast to general expectations, the electrical transport properties of the surface channels was found to be insensitive to high magnetic fields. Local and non-local magnetoresistance measurements allowed us to identify definite signatures of transport through multiple Dirac bands and strong inter-band scattering at the surface. Understanding the robustness of the surface states to magnetic fields requires further  experimental and theoretical investigations.

\begin{acknowledgments}
We acknowledge funding from Nanomission, Department of Science \& Technology (DST) and Indian Institute of Science. The work at the University of Warwick was supported by EPSRC, UK (Grant EP/L014963/1)
\end{acknowledgments}


\begin{thebibliography}{33}%
\makeatletter
\providecommand \@ifxundefined [1]{%
 \@ifx{#1\undefined}
}%
\providecommand \@ifnum [1]{%
 \ifnum #1\expandafter \@firstoftwo
 \else \expandafter \@secondoftwo
 \fi
}%
\providecommand \@ifx [1]{%
 \ifx #1\expandafter \@firstoftwo
 \else \expandafter \@secondoftwo
 \fi
}%
\providecommand \natexlab [1]{#1}%
\providecommand \enquote  [1]{``#1''}%
\providecommand \bibnamefont  [1]{#1}%
\providecommand \bibfnamefont [1]{#1}%
\providecommand \citenamefont [1]{#1}%
\providecommand \href@noop [0]{\@secondoftwo}%
\providecommand \href [0]{\begingroup \@sanitize@url \@href}%
\providecommand \@href[1]{\@@startlink{#1}\@@href}%
\providecommand \@@href[1]{\endgroup#1\@@endlink}%
\providecommand \@sanitize@url [0]{\catcode `\\12\catcode `\$12\catcode
  `\&12\catcode `\#12\catcode `\^12\catcode `\_12\catcode `\%12\relax}%
\providecommand \@@startlink[1]{}%
\providecommand \@@endlink[0]{}%
\providecommand \url  [0]{\begingroup\@sanitize@url \@url }%
\providecommand \@url [1]{\endgroup\@href {#1}{\urlprefix }}%
\providecommand \urlprefix  [0]{URL }%
\providecommand \Eprint [0]{\href }%
\providecommand \doibase [0]{http://dx.doi.org/}%
\providecommand \selectlanguage [0]{\@gobble}%
\providecommand \bibinfo  [0]{\@secondoftwo}%
\providecommand \bibfield  [0]{\@secondoftwo}%
\providecommand \translation [1]{[#1]}%
\providecommand \BibitemOpen [0]{}%
\providecommand \bibitemStop [0]{}%
\providecommand \bibitemNoStop [0]{.\EOS\space}%
\providecommand \EOS [0]{\spacefactor3000\relax}%
\providecommand \BibitemShut  [1]{\csname bibitem#1\endcsname}%
\let\auto@bib@innerbib\@empty
\bibitem [{\citenamefont {Hasan}\ and\ \citenamefont {Kane}(2010)}]{hassanrmp}%
  \BibitemOpen
  \bibfield  {author} {\bibinfo {author} {\bibfnamefont {M.~Z.}\ \bibnamefont
  {Hasan}}\ and\ \bibinfo {author} {\bibfnamefont {C.~L.}\ \bibnamefont
  {Kane}},\ }\href {\doibase 10.1103/RevModPhys.82.3045} {\bibfield  {journal}
  {\bibinfo  {journal} {Rev. Mod. Phys.}\ }\textbf {\bibinfo {volume} {82}},\
  \bibinfo {pages} {3045} (\bibinfo {year} {2010})}\BibitemShut {NoStop}%
\bibitem [{\citenamefont {Moore}(2010)}]{moore}%
  \BibitemOpen
  \bibfield  {author} {\bibinfo {author} {\bibfnamefont {J.~E.}\ \bibnamefont
  {Moore}},\ }\href@noop {} {\bibfield  {journal} {\bibinfo  {journal}
  {Nature}\ }\textbf {\bibinfo {volume} {464}},\ \bibinfo {pages} {194}
  (\bibinfo {year} {2010})}\BibitemShut {NoStop}%
\bibitem [{\citenamefont {Ando}(2013)}]{ando}%
  \BibitemOpen
  \bibfield  {author} {\bibinfo {author} {\bibfnamefont {Y.}~\bibnamefont
  {Ando}},\ }\href@noop {} {\bibfield  {journal} {\bibinfo  {journal} {Journal
  of the Physical Society of Japan}\ }\textbf {\bibinfo {volume} {82}},\
  \bibinfo {pages} {102001} (\bibinfo {year} {2013})}\BibitemShut {NoStop}%
\bibitem [{\citenamefont {Qi}\ and\ \citenamefont {Zhang}(2011)}]{zhangrmp}%
  \BibitemOpen
  \bibfield  {author} {\bibinfo {author} {\bibfnamefont {X.-L.}\ \bibnamefont
  {Qi}}\ and\ \bibinfo {author} {\bibfnamefont {S.-C.}\ \bibnamefont {Zhang}},\
  }\href {\doibase 10.1103/RevModPhys.83.1057} {\bibfield  {journal} {\bibinfo
  {journal} {Rev. Mod. Phys.}\ }\textbf {\bibinfo {volume} {83}},\ \bibinfo
  {pages} {1057} (\bibinfo {year} {2011})}\BibitemShut {NoStop}%
\bibitem [{\citenamefont {Kane}\ and\ \citenamefont
  {Mele}(2005)}]{PhysRevLett.95.226801}%
  \BibitemOpen
  \bibfield  {author} {\bibinfo {author} {\bibfnamefont {C.~L.}\ \bibnamefont
  {Kane}}\ and\ \bibinfo {author} {\bibfnamefont {E.~J.}\ \bibnamefont
  {Mele}},\ }\href {\doibase 10.1103/PhysRevLett.95.226801} {\bibfield
  {journal} {\bibinfo  {journal} {Phys. Rev. Lett.}\ }\textbf {\bibinfo
  {volume} {95}},\ \bibinfo {pages} {226801} (\bibinfo {year}
  {2005})}\BibitemShut {NoStop}%
\bibitem [{\citenamefont {Fu}\ and\ \citenamefont
  {Kane}(2008)}]{PhysRevLett.100.096407}%
  \BibitemOpen
  \bibfield  {author} {\bibinfo {author} {\bibfnamefont {L.}~\bibnamefont
  {Fu}}\ and\ \bibinfo {author} {\bibfnamefont {C.~L.}\ \bibnamefont {Kane}},\
  }\href {\doibase 10.1103/PhysRevLett.100.096407} {\bibfield  {journal}
  {\bibinfo  {journal} {Phys. Rev. Lett.}\ }\textbf {\bibinfo {volume} {100}},\
  \bibinfo {pages} {096407} (\bibinfo {year} {2008})}\BibitemShut {NoStop}%
\bibitem [{\citenamefont {Dzero}\ \emph {et~al.}(2010)\citenamefont {Dzero},
  \citenamefont {Sun}, \citenamefont {Galitski},\ and\ \citenamefont
  {Coleman}}]{dzero2010topological}%
  \BibitemOpen
  \bibfield  {author} {\bibinfo {author} {\bibfnamefont {M.}~\bibnamefont
  {Dzero}}, \bibinfo {author} {\bibfnamefont {K.}~\bibnamefont {Sun}}, \bibinfo
  {author} {\bibfnamefont {V.}~\bibnamefont {Galitski}}, \ and\ \bibinfo
  {author} {\bibfnamefont {P.}~\bibnamefont {Coleman}},\ }\href@noop {}
  {\bibfield  {journal} {\bibinfo  {journal} {Physical review letters}\
  }\textbf {\bibinfo {volume} {104}},\ \bibinfo {pages} {106408} (\bibinfo
  {year} {2010})}\BibitemShut {NoStop}%
\bibitem [{\citenamefont {Lu}\ \emph {et~al.}(2013)\citenamefont {Lu},
  \citenamefont {Zhao}, \citenamefont {Weng}, \citenamefont {Fang},\ and\
  \citenamefont {Dai}}]{lu2013correlated}%
  \BibitemOpen
  \bibfield  {author} {\bibinfo {author} {\bibfnamefont {F.}~\bibnamefont
  {Lu}}, \bibinfo {author} {\bibfnamefont {J.}~\bibnamefont {Zhao}}, \bibinfo
  {author} {\bibfnamefont {H.}~\bibnamefont {Weng}}, \bibinfo {author}
  {\bibfnamefont {Z.}~\bibnamefont {Fang}}, \ and\ \bibinfo {author}
  {\bibfnamefont {X.}~\bibnamefont {Dai}},\ }\href@noop {} {\bibfield
  {journal} {\bibinfo  {journal} {Physical review letters}\ }\textbf {\bibinfo
  {volume} {110}},\ \bibinfo {pages} {096401} (\bibinfo {year}
  {2013})}\BibitemShut {NoStop}%
\bibitem [{\citenamefont {Kim}\ \emph {et~al.}(2012)\citenamefont {Kim},
  \citenamefont {Grant},\ and\ \citenamefont {Fisk}}]{kim2012limit}%
  \BibitemOpen
  \bibfield  {author} {\bibinfo {author} {\bibfnamefont {D.}~\bibnamefont
  {Kim}}, \bibinfo {author} {\bibfnamefont {T.}~\bibnamefont {Grant}}, \ and\
  \bibinfo {author} {\bibfnamefont {Z.}~\bibnamefont {Fisk}},\ }\href@noop {}
  {\bibfield  {journal} {\bibinfo  {journal} {Physical review letters}\
  }\textbf {\bibinfo {volume} {109}},\ \bibinfo {pages} {096601} (\bibinfo
  {year} {2012})}\BibitemShut {NoStop}%
\bibitem [{\citenamefont {Zhang}\ \emph {et~al.}(2013)\citenamefont {Zhang},
  \citenamefont {Butch}, \citenamefont {Syers}, \citenamefont {Ziemak},
  \citenamefont {Greene},\ and\ \citenamefont
  {Paglione}}]{zhang2013hybridization}%
  \BibitemOpen
  \bibfield  {author} {\bibinfo {author} {\bibfnamefont {X.}~\bibnamefont
  {Zhang}}, \bibinfo {author} {\bibfnamefont {N.}~\bibnamefont {Butch}},
  \bibinfo {author} {\bibfnamefont {P.}~\bibnamefont {Syers}}, \bibinfo
  {author} {\bibfnamefont {S.}~\bibnamefont {Ziemak}}, \bibinfo {author}
  {\bibfnamefont {R.~L.}\ \bibnamefont {Greene}}, \ and\ \bibinfo {author}
  {\bibfnamefont {J.}~\bibnamefont {Paglione}},\ }\href@noop {} {\bibfield
  {journal} {\bibinfo  {journal} {Physical Review X}\ }\textbf {\bibinfo
  {volume} {3}},\ \bibinfo {pages} {011011} (\bibinfo {year}
  {2013})}\BibitemShut {NoStop}%
\bibitem [{\citenamefont {Kim}\ \emph {et~al.}(2014)\citenamefont {Kim},
  \citenamefont {Xia},\ and\ \citenamefont {Fisk}}]{kim2014topological}%
  \BibitemOpen
  \bibfield  {author} {\bibinfo {author} {\bibfnamefont {D.-J.}\ \bibnamefont
  {Kim}}, \bibinfo {author} {\bibfnamefont {J.}~\bibnamefont {Xia}}, \ and\
  \bibinfo {author} {\bibfnamefont {Z.}~\bibnamefont {Fisk}},\ }\href@noop {}
  {\bibfield  {journal} {\bibinfo  {journal} {Nature materials}\ }\textbf
  {\bibinfo {volume} {13}},\ \bibinfo {pages} {466} (\bibinfo {year}
  {2014})}\BibitemShut {NoStop}%
\bibitem [{\citenamefont {Kim}\ \emph {et~al.}(2013)\citenamefont {Kim},
  \citenamefont {Thomas}, \citenamefont {Grant}, \citenamefont {Botimer},
  \citenamefont {Fisk},\ and\ \citenamefont {Xia}}]{kim2013surface}%
  \BibitemOpen
  \bibfield  {author} {\bibinfo {author} {\bibfnamefont {D.}~\bibnamefont
  {Kim}}, \bibinfo {author} {\bibfnamefont {S.}~\bibnamefont {Thomas}},
  \bibinfo {author} {\bibfnamefont {T.}~\bibnamefont {Grant}}, \bibinfo
  {author} {\bibfnamefont {J.}~\bibnamefont {Botimer}}, \bibinfo {author}
  {\bibfnamefont {Z.}~\bibnamefont {Fisk}}, \ and\ \bibinfo {author}
  {\bibfnamefont {J.}~\bibnamefont {Xia}},\ }\href@noop {} {\bibfield
  {journal} {\bibinfo  {journal} {Scientific reports}\ }\textbf {\bibinfo
  {volume} {3}} (\bibinfo {year} {2013})}\BibitemShut {NoStop}%
\bibitem [{\citenamefont {Wolgast}\ \emph {et~al.}(2013)\citenamefont
  {Wolgast}, \citenamefont {Kurdak}, \citenamefont {Sun}, \citenamefont
  {Allen}, \citenamefont {Kim},\ and\ \citenamefont {Fisk}}]{wolgast2013low}%
  \BibitemOpen
  \bibfield  {author} {\bibinfo {author} {\bibfnamefont {S.}~\bibnamefont
  {Wolgast}}, \bibinfo {author} {\bibfnamefont {{\c{C}}.}~\bibnamefont
  {Kurdak}}, \bibinfo {author} {\bibfnamefont {K.}~\bibnamefont {Sun}},
  \bibinfo {author} {\bibfnamefont {J.}~\bibnamefont {Allen}}, \bibinfo
  {author} {\bibfnamefont {D.-J.}\ \bibnamefont {Kim}}, \ and\ \bibinfo
  {author} {\bibfnamefont {Z.}~\bibnamefont {Fisk}},\ }\href@noop {} {\bibfield
   {journal} {\bibinfo  {journal} {Physical Review B}\ }\textbf {\bibinfo
  {volume} {88}},\ \bibinfo {pages} {180405} (\bibinfo {year}
  {2013})}\BibitemShut {NoStop}%
\bibitem [{\citenamefont {Jiang}\ \emph {et~al.}(2013)\citenamefont {Jiang},
  \citenamefont {Li}, \citenamefont {Zhang}, \citenamefont {Sun}, \citenamefont
  {Chen}, \citenamefont {Ye}, \citenamefont {Xu}, \citenamefont {Ge},
  \citenamefont {Tan}, \citenamefont {Niu} \emph
  {et~al.}}]{jiang2013observation}%
  \BibitemOpen
  \bibfield  {author} {\bibinfo {author} {\bibfnamefont {J.}~\bibnamefont
  {Jiang}}, \bibinfo {author} {\bibfnamefont {S.}~\bibnamefont {Li}}, \bibinfo
  {author} {\bibfnamefont {T.}~\bibnamefont {Zhang}}, \bibinfo {author}
  {\bibfnamefont {Z.}~\bibnamefont {Sun}}, \bibinfo {author} {\bibfnamefont
  {F.}~\bibnamefont {Chen}}, \bibinfo {author} {\bibfnamefont {Z.}~\bibnamefont
  {Ye}}, \bibinfo {author} {\bibfnamefont {M.}~\bibnamefont {Xu}}, \bibinfo
  {author} {\bibfnamefont {Q.}~\bibnamefont {Ge}}, \bibinfo {author}
  {\bibfnamefont {S.}~\bibnamefont {Tan}}, \bibinfo {author} {\bibfnamefont
  {X.}~\bibnamefont {Niu}},  \emph {et~al.},\ }\href@noop {} {\bibfield
  {journal} {\bibinfo  {journal} {Nature communications}\ }\textbf {\bibinfo
  {volume} {4}} (\bibinfo {year} {2013})}\BibitemShut {NoStop}%
\bibitem [{\citenamefont {Xu}\ \emph {et~al.}(2014{\natexlab{a}})\citenamefont
  {Xu}, \citenamefont {Matt}, \citenamefont {Pomjakushina}, \citenamefont
  {Shi}, \citenamefont {Dhaka}, \citenamefont {Plumb}, \citenamefont
  {Radovi{\'c}}, \citenamefont {Biswas}, \citenamefont {Evtushinsky},
  \citenamefont {Zabolotnyy} \emph {et~al.}}]{xu2014exotic}%
  \BibitemOpen
  \bibfield  {author} {\bibinfo {author} {\bibfnamefont {N.}~\bibnamefont
  {Xu}}, \bibinfo {author} {\bibfnamefont {C.}~\bibnamefont {Matt}}, \bibinfo
  {author} {\bibfnamefont {E.}~\bibnamefont {Pomjakushina}}, \bibinfo {author}
  {\bibfnamefont {X.}~\bibnamefont {Shi}}, \bibinfo {author} {\bibfnamefont
  {R.}~\bibnamefont {Dhaka}}, \bibinfo {author} {\bibfnamefont
  {N.}~\bibnamefont {Plumb}}, \bibinfo {author} {\bibfnamefont
  {M.}~\bibnamefont {Radovi{\'c}}}, \bibinfo {author} {\bibfnamefont
  {P.}~\bibnamefont {Biswas}}, \bibinfo {author} {\bibfnamefont
  {D.}~\bibnamefont {Evtushinsky}}, \bibinfo {author} {\bibfnamefont
  {V.}~\bibnamefont {Zabolotnyy}},  \emph {et~al.},\ }\href@noop {} {\bibfield
  {journal} {\bibinfo  {journal} {Physical Review B}\ }\textbf {\bibinfo
  {volume} {90}},\ \bibinfo {pages} {085148} (\bibinfo {year}
  {2014}{\natexlab{a}})}\BibitemShut {NoStop}%
\bibitem [{\citenamefont {Xu}\ \emph {et~al.}(2014{\natexlab{b}})\citenamefont
  {Xu}, \citenamefont {Biswas}, \citenamefont {Dil}, \citenamefont {Dhaka},
  \citenamefont {Landolt}, \citenamefont {Muff}, \citenamefont {Matt},
  \citenamefont {Shi}, \citenamefont {Plumb}, \citenamefont {Radovi{\'c}} \emph
  {et~al.}}]{xu2014direct}%
  \BibitemOpen
  \bibfield  {author} {\bibinfo {author} {\bibfnamefont {N.}~\bibnamefont
  {Xu}}, \bibinfo {author} {\bibfnamefont {P.}~\bibnamefont {Biswas}}, \bibinfo
  {author} {\bibfnamefont {J.}~\bibnamefont {Dil}}, \bibinfo {author}
  {\bibfnamefont {R.}~\bibnamefont {Dhaka}}, \bibinfo {author} {\bibfnamefont
  {G.}~\bibnamefont {Landolt}}, \bibinfo {author} {\bibfnamefont
  {S.}~\bibnamefont {Muff}}, \bibinfo {author} {\bibfnamefont {C.}~\bibnamefont
  {Matt}}, \bibinfo {author} {\bibfnamefont {X.}~\bibnamefont {Shi}}, \bibinfo
  {author} {\bibfnamefont {N.}~\bibnamefont {Plumb}}, \bibinfo {author}
  {\bibfnamefont {M.}~\bibnamefont {Radovi{\'c}}},  \emph {et~al.},\
  }\href@noop {} {\bibfield  {journal} {\bibinfo  {journal} {Nature
  communications}\ }\textbf {\bibinfo {volume} {5}} (\bibinfo {year}
  {2014}{\natexlab{b}})}\BibitemShut {NoStop}%
\bibitem [{\citenamefont {Liu}\ \emph {et~al.}(2012)\citenamefont {Liu},
  \citenamefont {Zhang}, \citenamefont {Chang}, \citenamefont {Zhang},
  \citenamefont {Feng}, \citenamefont {Li}, \citenamefont {He}, \citenamefont
  {Wang}, \citenamefont {Chen}, \citenamefont {Dai}, \citenamefont {Fang},
  \citenamefont {Xue}, \citenamefont {Ma},\ and\ \citenamefont
  {Wang}}]{PhysRevLett.108.036805}%
  \BibitemOpen
  \bibfield  {author} {\bibinfo {author} {\bibfnamefont {M.}~\bibnamefont
  {Liu}}, \bibinfo {author} {\bibfnamefont {J.}~\bibnamefont {Zhang}}, \bibinfo
  {author} {\bibfnamefont {C.-Z.}\ \bibnamefont {Chang}}, \bibinfo {author}
  {\bibfnamefont {Z.}~\bibnamefont {Zhang}}, \bibinfo {author} {\bibfnamefont
  {X.}~\bibnamefont {Feng}}, \bibinfo {author} {\bibfnamefont {K.}~\bibnamefont
  {Li}}, \bibinfo {author} {\bibfnamefont {K.}~\bibnamefont {He}}, \bibinfo
  {author} {\bibfnamefont {L.-l.}\ \bibnamefont {Wang}}, \bibinfo {author}
  {\bibfnamefont {X.}~\bibnamefont {Chen}}, \bibinfo {author} {\bibfnamefont
  {X.}~\bibnamefont {Dai}}, \bibinfo {author} {\bibfnamefont {Z.}~\bibnamefont
  {Fang}}, \bibinfo {author} {\bibfnamefont {Q.-K.}\ \bibnamefont {Xue}},
  \bibinfo {author} {\bibfnamefont {X.}~\bibnamefont {Ma}}, \ and\ \bibinfo
  {author} {\bibfnamefont {Y.}~\bibnamefont {Wang}},\ }\href {\doibase
  10.1103/PhysRevLett.108.036805} {\bibfield  {journal} {\bibinfo  {journal}
  {Phys. Rev. Lett.}\ }\textbf {\bibinfo {volume} {108}},\ \bibinfo {pages}
  {036805} (\bibinfo {year} {2012})}\BibitemShut {NoStop}%
\bibitem [{\citenamefont {Shekhar}\ \emph {et~al.}(2014)\citenamefont
  {Shekhar}, \citenamefont {ViolBarbosa}, \citenamefont {Yan}, \citenamefont
  {Ouardi}, \citenamefont {Schnelle}, \citenamefont {Fecher},\ and\
  \citenamefont {Felser}}]{PhysRevB.90.165140}%
  \BibitemOpen
  \bibfield  {author} {\bibinfo {author} {\bibfnamefont {C.}~\bibnamefont
  {Shekhar}}, \bibinfo {author} {\bibfnamefont {C.~E.}\ \bibnamefont
  {ViolBarbosa}}, \bibinfo {author} {\bibfnamefont {B.}~\bibnamefont {Yan}},
  \bibinfo {author} {\bibfnamefont {S.}~\bibnamefont {Ouardi}}, \bibinfo
  {author} {\bibfnamefont {W.}~\bibnamefont {Schnelle}}, \bibinfo {author}
  {\bibfnamefont {G.~H.}\ \bibnamefont {Fecher}}, \ and\ \bibinfo {author}
  {\bibfnamefont {C.}~\bibnamefont {Felser}},\ }\href {\doibase
  10.1103/PhysRevB.90.165140} {\bibfield  {journal} {\bibinfo  {journal} {Phys.
  Rev. B}\ }\textbf {\bibinfo {volume} {90}},\ \bibinfo {pages} {165140}
  (\bibinfo {year} {2014})}\BibitemShut {NoStop}%
\bibitem [{\citenamefont {He}\ \emph {et~al.}(2011)\citenamefont {He},
  \citenamefont {Wang}, \citenamefont {Zhang}, \citenamefont {Sou},
  \citenamefont {Wong}, \citenamefont {Wang}, \citenamefont {Lu}, \citenamefont
  {Shen},\ and\ \citenamefont {Zhang}}]{PhysRevLett.106.166805}%
  \BibitemOpen
  \bibfield  {author} {\bibinfo {author} {\bibfnamefont {H.-T.}\ \bibnamefont
  {He}}, \bibinfo {author} {\bibfnamefont {G.}~\bibnamefont {Wang}}, \bibinfo
  {author} {\bibfnamefont {T.}~\bibnamefont {Zhang}}, \bibinfo {author}
  {\bibfnamefont {I.-K.}\ \bibnamefont {Sou}}, \bibinfo {author} {\bibfnamefont
  {G.~K.~L.}\ \bibnamefont {Wong}}, \bibinfo {author} {\bibfnamefont {J.-N.}\
  \bibnamefont {Wang}}, \bibinfo {author} {\bibfnamefont {H.-Z.}\ \bibnamefont
  {Lu}}, \bibinfo {author} {\bibfnamefont {S.-Q.}\ \bibnamefont {Shen}}, \ and\
  \bibinfo {author} {\bibfnamefont {F.-C.}\ \bibnamefont {Zhang}},\ }\href
  {\doibase 10.1103/PhysRevLett.106.166805} {\bibfield  {journal} {\bibinfo
  {journal} {Phys. Rev. Lett.}\ }\textbf {\bibinfo {volume} {106}},\ \bibinfo
  {pages} {166805} (\bibinfo {year} {2011})}\BibitemShut {NoStop}%
\bibitem [{\citenamefont {{Thomas}}\ \emph {et~al.}(2013)\citenamefont
  {{Thomas}}, \citenamefont {{Kim}}, \citenamefont {{Chung}}, \citenamefont
  {{Grant}}, \citenamefont {{Fisk}},\ and\ \citenamefont
  {{Xia}}}]{2013arXiv1307.4133T}%
  \BibitemOpen
  \bibfield  {author} {\bibinfo {author} {\bibfnamefont {S.}~\bibnamefont
  {{Thomas}}}, \bibinfo {author} {\bibfnamefont {D.~J.}\ \bibnamefont {{Kim}}},
  \bibinfo {author} {\bibfnamefont {S.~B.}\ \bibnamefont {{Chung}}}, \bibinfo
  {author} {\bibfnamefont {T.}~\bibnamefont {{Grant}}}, \bibinfo {author}
  {\bibfnamefont {Z.}~\bibnamefont {{Fisk}}}, \ and\ \bibinfo {author}
  {\bibfnamefont {J.}~\bibnamefont {{Xia}}},\ }\href@noop {} {\bibfield
  {journal} {\bibinfo  {journal} {ArXiv e-prints}\ } (\bibinfo {year}
  {2013})},\ \Eprint {http://arxiv.org/abs/1307.4133} {arXiv:1307.4133
  [cond-mat.str-el]} \BibitemShut {NoStop}%
\bibitem [{\citenamefont {Ciomaga~Hatnean}\ \emph {et~al.}(2013)\citenamefont
  {Ciomaga~Hatnean}, \citenamefont {Lees}, \citenamefont {Paul},\ and\
  \citenamefont {Balakrishnan}}]{balakrishnan}%
  \BibitemOpen
  \bibfield  {author} {\bibinfo {author} {\bibfnamefont {M.}~\bibnamefont
  {Ciomaga~Hatnean}}, \bibinfo {author} {\bibfnamefont {M.~R.}\ \bibnamefont
  {Lees}}, \bibinfo {author} {\bibfnamefont {D.~M.}\ \bibnamefont {Paul}}, \
  and\ \bibinfo {author} {\bibfnamefont {G.}~\bibnamefont {Balakrishnan}},\
  }\href@noop {} {\bibfield  {journal} {\bibinfo  {journal} {Sci. Rep.}\
  }\textbf {\bibinfo {volume} {3}},\ \bibinfo {pages} {1} (\bibinfo {year}
  {2013})}\BibitemShut {NoStop}%
\bibitem [{\citenamefont {Allen}\ \emph {et~al.}(1979)\citenamefont {Allen},
  \citenamefont {Batlogg},\ and\ \citenamefont {Wachter}}]{allen1979large}%
  \BibitemOpen
  \bibfield  {author} {\bibinfo {author} {\bibfnamefont {J.}~\bibnamefont
  {Allen}}, \bibinfo {author} {\bibfnamefont {B.}~\bibnamefont {Batlogg}}, \
  and\ \bibinfo {author} {\bibfnamefont {P.}~\bibnamefont {Wachter}},\
  }\href@noop {} {\bibfield  {journal} {\bibinfo  {journal} {Physical Review
  B}\ }\textbf {\bibinfo {volume} {20}},\ \bibinfo {pages} {4807} (\bibinfo
  {year} {1979})}\BibitemShut {NoStop}%
\bibitem [{\citenamefont {http://www.comsol.com}()}]{comsol}%
  \BibitemOpen
  \bibfield  {author} {\bibinfo {author} {\bibnamefont
  {http://www.comsol.com}},\ }\href@noop {} {\ }\BibitemShut {NoStop}%
\bibitem [{\citenamefont {Sheng}\ \emph {et~al.}(2005)\citenamefont {Sheng},
  \citenamefont {Sheng}, \citenamefont {Ting},\ and\ \citenamefont
  {Haldane}}]{PhysRevLett.95.136602}%
  \BibitemOpen
  \bibfield  {author} {\bibinfo {author} {\bibfnamefont {L.}~\bibnamefont
  {Sheng}}, \bibinfo {author} {\bibfnamefont {D.~N.}\ \bibnamefont {Sheng}},
  \bibinfo {author} {\bibfnamefont {C.~S.}\ \bibnamefont {Ting}}, \ and\
  \bibinfo {author} {\bibfnamefont {F.~D.~M.}\ \bibnamefont {Haldane}},\ }\href
  {\doibase 10.1103/PhysRevLett.95.136602} {\bibfield  {journal} {\bibinfo
  {journal} {Phys. Rev. Lett.}\ }\textbf {\bibinfo {volume} {95}},\ \bibinfo
  {pages} {136602} (\bibinfo {year} {2005})}\BibitemShut {NoStop}%
\bibitem [{\citenamefont {Sheng}\ \emph {et~al.}(2006)\citenamefont {Sheng},
  \citenamefont {Weng}, \citenamefont {Sheng},\ and\ \citenamefont
  {Haldane}}]{PhysRevLett.97.036808}%
  \BibitemOpen
  \bibfield  {author} {\bibinfo {author} {\bibfnamefont {D.~N.}\ \bibnamefont
  {Sheng}}, \bibinfo {author} {\bibfnamefont {Z.~Y.}\ \bibnamefont {Weng}},
  \bibinfo {author} {\bibfnamefont {L.}~\bibnamefont {Sheng}}, \ and\ \bibinfo
  {author} {\bibfnamefont {F.~D.~M.}\ \bibnamefont {Haldane}},\ }\href
  {\doibase 10.1103/PhysRevLett.97.036808} {\bibfield  {journal} {\bibinfo
  {journal} {Phys. Rev. Lett.}\ }\textbf {\bibinfo {volume} {97}},\ \bibinfo
  {pages} {036808} (\bibinfo {year} {2006})}\BibitemShut {NoStop}%
\bibitem [{\citenamefont {Prodan}(2009)}]{PhysRevB.80.125327}%
  \BibitemOpen
  \bibfield  {author} {\bibinfo {author} {\bibfnamefont {E.}~\bibnamefont
  {Prodan}},\ }\href {\doibase 10.1103/PhysRevB.80.125327} {\bibfield
  {journal} {\bibinfo  {journal} {Phys. Rev. B}\ }\textbf {\bibinfo {volume}
  {80}},\ \bibinfo {pages} {125327} (\bibinfo {year} {2009})}\BibitemShut
  {NoStop}%
\bibitem [{\citenamefont {Yang}\ \emph {et~al.}(2011)\citenamefont {Yang},
  \citenamefont {Xu}, \citenamefont {Sheng}, \citenamefont {Wang},
  \citenamefont {Xing},\ and\ \citenamefont {Sheng}}]{PhysRevLett.107.066602}%
  \BibitemOpen
  \bibfield  {author} {\bibinfo {author} {\bibfnamefont {Y.}~\bibnamefont
  {Yang}}, \bibinfo {author} {\bibfnamefont {Z.}~\bibnamefont {Xu}}, \bibinfo
  {author} {\bibfnamefont {L.}~\bibnamefont {Sheng}}, \bibinfo {author}
  {\bibfnamefont {B.}~\bibnamefont {Wang}}, \bibinfo {author} {\bibfnamefont
  {D.~Y.}\ \bibnamefont {Xing}}, \ and\ \bibinfo {author} {\bibfnamefont
  {D.~N.}\ \bibnamefont {Sheng}},\ }\href {\doibase
  10.1103/PhysRevLett.107.066602} {\bibfield  {journal} {\bibinfo  {journal}
  {Phys. Rev. Lett.}\ }\textbf {\bibinfo {volume} {107}},\ \bibinfo {pages}
  {066602} (\bibinfo {year} {2011})}\BibitemShut {NoStop}%
\bibitem [{\citenamefont {Du}\ \emph {et~al.}(2013)\citenamefont {Du},
  \citenamefont {Knez}, \citenamefont {Sullivan},\ and\ \citenamefont
  {Du}}]{du2013observation}%
  \BibitemOpen
  \bibfield  {author} {\bibinfo {author} {\bibfnamefont {L.}~\bibnamefont
  {Du}}, \bibinfo {author} {\bibfnamefont {I.}~\bibnamefont {Knez}}, \bibinfo
  {author} {\bibfnamefont {G.}~\bibnamefont {Sullivan}}, \ and\ \bibinfo
  {author} {\bibfnamefont {R.-R.}\ \bibnamefont {Du}},\ }\href@noop {}
  {\bibfield  {journal} {\bibinfo  {journal} {arXiv preprint arXiv:1306.1925}\
  } (\bibinfo {year} {2013})}\BibitemShut {NoStop}%
\bibitem [{\citenamefont {Bergmann}(1983)}]{PhysRevB.28.2914}%
  \BibitemOpen
  \bibfield  {author} {\bibinfo {author} {\bibfnamefont {G.}~\bibnamefont
  {Bergmann}},\ }\href {\doibase 10.1103/PhysRevB.28.2914} {\bibfield
  {journal} {\bibinfo  {journal} {Phys. Rev. B}\ }\textbf {\bibinfo {volume}
  {28}},\ \bibinfo {pages} {2914} (\bibinfo {year} {1983})}\BibitemShut
  {NoStop}%
\bibitem [{\citenamefont {Pierre}\ \emph {et~al.}(2003)\citenamefont {Pierre},
  \citenamefont {Gougam}, \citenamefont {Anthore}, \citenamefont {Pothier},
  \citenamefont {Esteve},\ and\ \citenamefont {Birge}}]{PhysRevB.68.085413}%
  \BibitemOpen
  \bibfield  {author} {\bibinfo {author} {\bibfnamefont {F.}~\bibnamefont
  {Pierre}}, \bibinfo {author} {\bibfnamefont {A.~B.}\ \bibnamefont {Gougam}},
  \bibinfo {author} {\bibfnamefont {A.}~\bibnamefont {Anthore}}, \bibinfo
  {author} {\bibfnamefont {H.}~\bibnamefont {Pothier}}, \bibinfo {author}
  {\bibfnamefont {D.}~\bibnamefont {Esteve}}, \ and\ \bibinfo {author}
  {\bibfnamefont {N.~O.}\ \bibnamefont {Birge}},\ }\href {\doibase
  10.1103/PhysRevB.68.085413} {\bibfield  {journal} {\bibinfo  {journal} {Phys.
  Rev. B}\ }\textbf {\bibinfo {volume} {68}},\ \bibinfo {pages} {085413}
  (\bibinfo {year} {2003})}\BibitemShut {NoStop}%
\bibitem [{\citenamefont {Hikami}\ \emph {et~al.}(1980)\citenamefont {Hikami},
  \citenamefont {Larkin},\ and\ \citenamefont {Nagaoka}}]{hikami1980spin}%
  \BibitemOpen
  \bibfield  {author} {\bibinfo {author} {\bibfnamefont {S.}~\bibnamefont
  {Hikami}}, \bibinfo {author} {\bibfnamefont {A.~I.}\ \bibnamefont {Larkin}},
  \ and\ \bibinfo {author} {\bibfnamefont {Y.}~\bibnamefont {Nagaoka}},\
  }\href@noop {} {\bibfield  {journal} {\bibinfo  {journal} {Progress of
  Theoretical Physics}\ }\textbf {\bibinfo {volume} {63}},\ \bibinfo {pages}
  {707} (\bibinfo {year} {1980})}\BibitemShut {NoStop}%
\bibitem [{\citenamefont {Chiu}\ and\ \citenamefont
  {Lin}(2013)}]{PhysRevB.87.035122}%
  \BibitemOpen
  \bibfield  {author} {\bibinfo {author} {\bibfnamefont {S.-P.}\ \bibnamefont
  {Chiu}}\ and\ \bibinfo {author} {\bibfnamefont {J.-J.}\ \bibnamefont {Lin}},\
  }\href {\doibase 10.1103/PhysRevB.87.035122} {\bibfield  {journal} {\bibinfo
  {journal} {Phys. Rev. B}\ }\textbf {\bibinfo {volume} {87}},\ \bibinfo
  {pages} {035122} (\bibinfo {year} {2013})}\BibitemShut {NoStop}%
\bibitem [{\citenamefont {Lee}\ \emph {et~al.}(2014)\citenamefont {Lee},
  \citenamefont {Lee}, \citenamefont {Park}, \citenamefont {Kim},\ and\
  \citenamefont {Lee}}]{PhysRevX.4.011039}%
  \BibitemOpen
  \bibfield  {author} {\bibinfo {author} {\bibfnamefont {J.}~\bibnamefont
  {Lee}}, \bibinfo {author} {\bibfnamefont {J.-H.}\ \bibnamefont {Lee}},
  \bibinfo {author} {\bibfnamefont {J.}~\bibnamefont {Park}}, \bibinfo {author}
  {\bibfnamefont {J.~S.}\ \bibnamefont {Kim}}, \ and\ \bibinfo {author}
  {\bibfnamefont {H.-J.}\ \bibnamefont {Lee}},\ }\href {\doibase
  10.1103/PhysRevX.4.011039} {\bibfield  {journal} {\bibinfo  {journal} {Phys.
  Rev. X}\ }\textbf {\bibinfo {volume} {4}},\ \bibinfo {pages} {011039}
  (\bibinfo {year} {2014})}\BibitemShut {NoStop}%
\end{thebibliography}

%

\pagebreak

\section*{Supplementary Material}

\begin{table*}[!]
\begin{ruledtabular}
\begin{tabular}{| l | c | c | c | c | c | c |}
Configuration & Source & Drain & Local V$_+$ & Local V$_-$ & Non-local V$_+$ & Non-local V$_-$ \\ \hline
A &4&7&5&6&3&2\\ \hline
C &1&4&2&3&6&5\\ \hline
E &7&10&8&9&6&5\\ \hline
1NL &2&1&variable&variable&variable&variable\\ \hline
2NL &9&10&variable&variable&variable&variable\\ \hline
\end{tabular}
\small{\caption{Description of the electrical contact configurations used in this work, the numbers refer to the elecrical probes as shown in figure\ref{fig:rt}(a). In the case of configurations 1NL and 2NL, the voltage probes were varied depending on the measurement.  \label{tab:config}}}
 
\end{ruledtabular}
\end{table*}

\section*{ Calculation of surface and bulk current components} 

\begin{figure}[tbh]
\begin{center}
\includegraphics[width=0.45\textwidth]{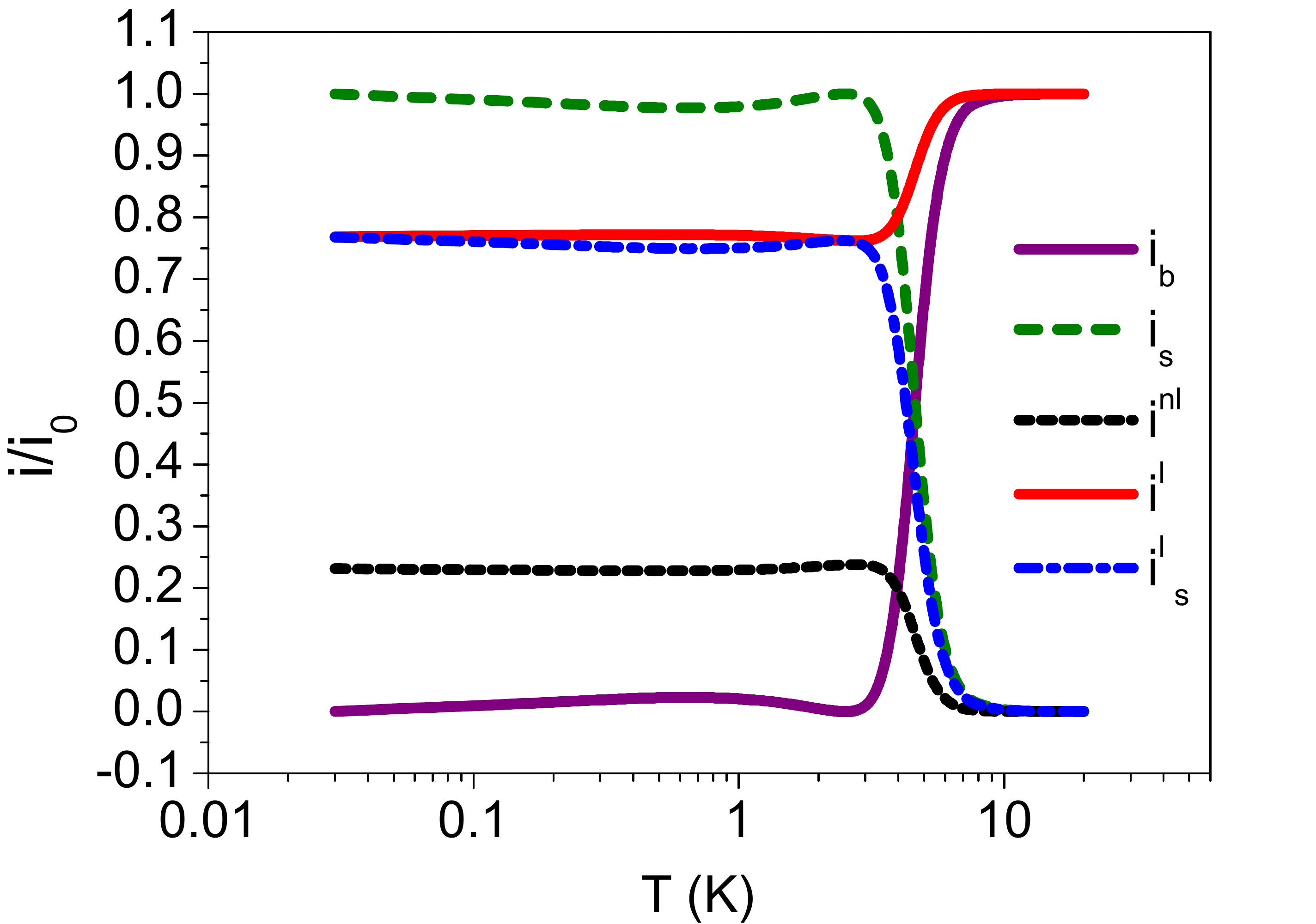}
\small{\caption{Different current components flowing through the device as a function of temperature as deduced from our analysis. The values have been normalized by the bias current $i_0$. \label{fig:currents}}}
\end{center}
\end{figure}

At ultra-low temperatures when the bulk conductance has frozen out, the applied current flows entirely through the surface, i.e. $i_s=i_0$. Then, we can write
\begin{eqnarray}
V^l(T)=i^{l}_{s}(T) R_s(T); V^{nl}(T) = i^{nl}_{s}(T) R_s(T)
\end{eqnarray}
From these two equations and  $i^{l}_{s} = \gamma i_{s} = \gamma i_0 $ we get, $\gamma$=(V$^{l}_{m}$)/(V$^{l}_{m}$+V$^{nl}_{m}$) and $R_s$ = ((V$^{l}_{m}$+V$^{nl}_{m}$)/$i_0$. From the measured values of \vnl and \vl we get $\gamma$= 0.87 , 0.77 and 0.81 for configuration A, C and E respectively (see Table 1).~\ref{tab:config} Assuming that the change in the surface resistance  $R_s$ with temperature is negligible in comparison to the orders of magnitude change in $R_b$, using $i^{nl}_{s}(T)=V_{nl}(T)/R_s$), we get the value of the temperature dependent non-local surface current. Subtracting this from the constant total current $i_0$ supplied to the sample gives the temperature dependent local current $i^{l}(T)$ flowing through the sample. Subtracting the local surface current $i^{l}_{s}(T) (=V_l(T)/R_s$) from the total local current $i^{l}(T)$ gives us bulk current $i_{b}(T)$. The values of the different current components obtained through this analysis are plotted in figure~\ref{fig:currents} as a function of temperature. It can be seen that at high temperatures the entire current flows through the bulk of the device. At temperatures below 3~K $i_b$ drops exponentially and the entire current shifts to the surface channels. The non-local current $i^{nl}$, which was negligible till now, becomes a significant fraction (almost 20$\%$) of the total current. The rest of the current flows through the local surface channel.

\begin{figure}[tbh]
\begin{center}
\includegraphics[width=0.45\textwidth]{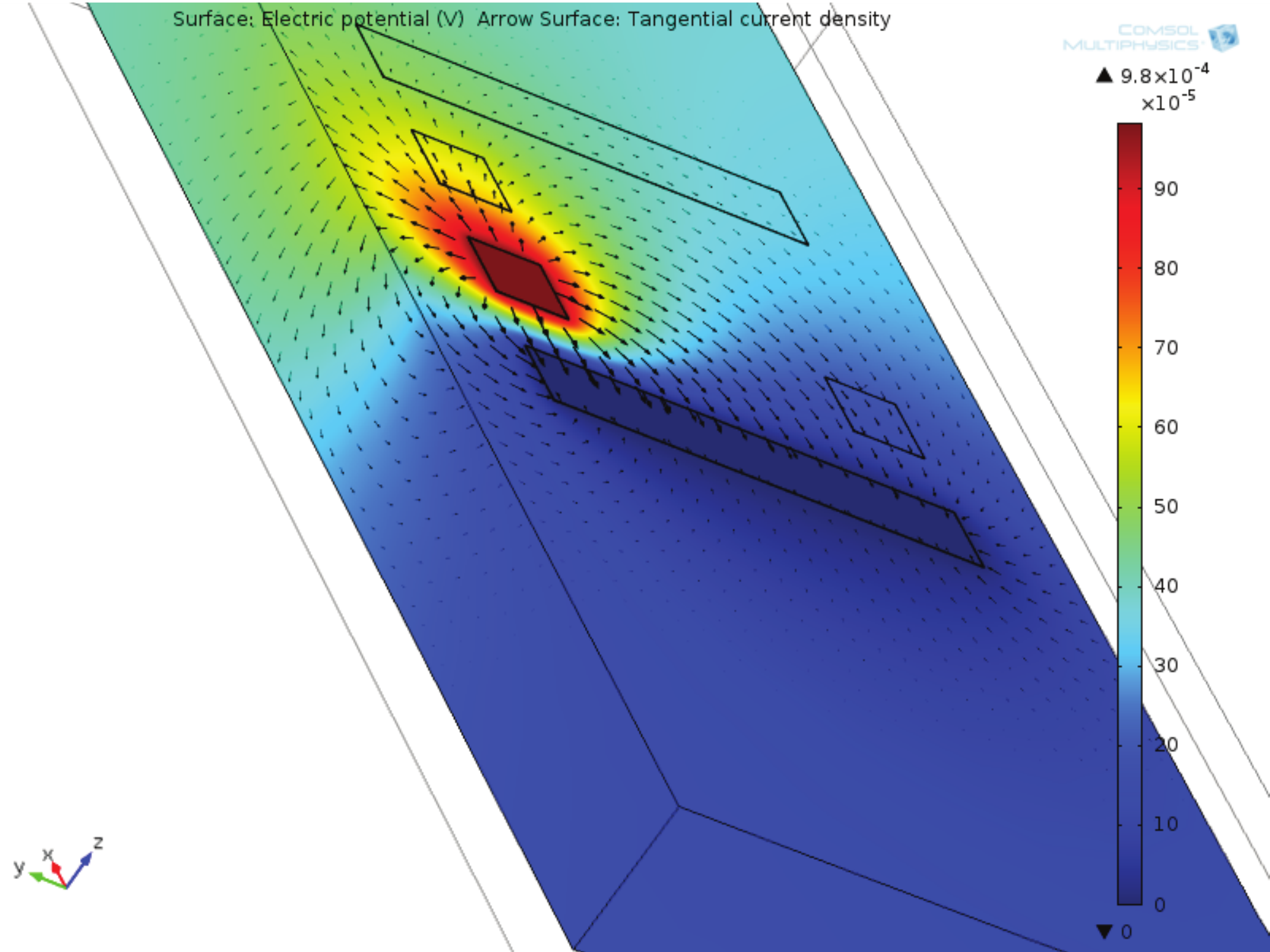}
\small{\caption{ Electric field profile obtained from our simulations.\label{fig:sim_image}}}
\end{center}
\end{figure}

The geometry dependent factor $\gamma$ is decided by the effective resistances of the local and non-local transport channels 
\begin{eqnarray}
\gamma = R^{nl}/(R^{nl}+R^{l})
\end{eqnarray}
which in turn depend on the lengths and surface resistances of the local and non-local transport paths. $R^l$ depends only on the properties of the top surface as the local current flows only on the top surface. $R^{nl}$ on the other hand depends on the properties of the top, bottom and the side surfaces. Assuming these surfaces have very similar specific resistances, we estimate $\gamma$=0.945 for our device. Our experiments found the $\gamma$ value to be almost $15\%$ lower than this value. We believe this to be due to two reasons. Firstly, our simulations show that there is a  significant fringing of the electric field lines implying that the non-local current flows through all six sides of the device. The second reason is that the transport in this material is highly anisotropic and the assumption that all the surfaces have similar specific resistance is strictly not valid.

To understand the experimentally obtained value of $\gamma$ better, we performed extensive numerical simulations. The simulations reproduce the experimentally measured value of \vnl and their dependence on the distance from source contact $d$ as can be seen in figure~\ref{fig:nlsim}.

\section*{Details of simulation} 

\begin{figure}[tbh]
\begin{center}
\includegraphics[width=0.45\textwidth]{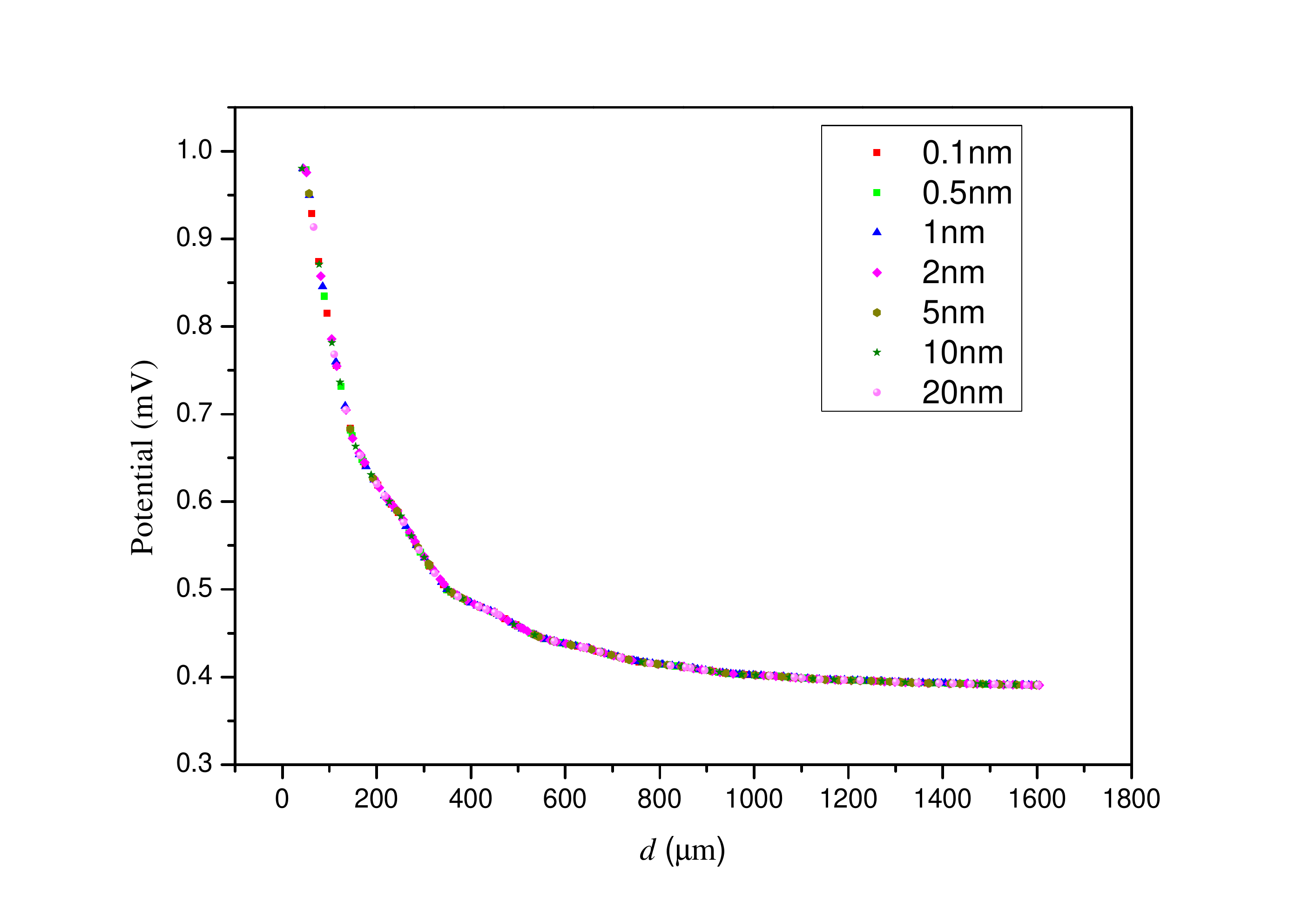}
\small{\caption{Calculated \vnl as a function of distance of the voltage probes from the source electrode, the numbers in the legend refer to the effective thickness of the conducting layer used for the simulations. It is seen that the various data sets collapse on top of each other.  \label{fig:sim_data}}}
\end{center}
\end{figure}

The system was simulated using COMSOL Multiphysics modelling software~\cite{comsol}. The geometry of the sample was designed with the actual dimensions. The bulk was assumed to be an ideal insulator while the surface was taken to be a conducting channel of thickness $t$ with resistance $R_s = 294 \Omega/\Box$. The numerical simulation involved the solution of the continuity equation with the bias applied as the boundary condition to obtain the potential map~\ref{fig:sim_image}.This was done for the various configurations listed in table 1~\label{tab:config}. The effective thickness of the conducting channel was taken to be 1 \AA. This is justified since, it was observed that the result of simulation did not change by varying the effective channel thickness in the range $t$ = 1-200 \AA (see figure~\ref{fig:sim_data}) and the spatial distribution of the surface state has been reported to be around 10-20 \AA~\cite{PhysRevX.4.011039}.

\end{document}